\documentclass[prb, twocolumn, 10pt]{revtex4}
\usepackage{graphicx}
\usepackage{dcolumn}
\usepackage{bm}
\begin{document}
\title{X-ray absorption study of Ti-activated sodium aluminum hydride}
\author{J. Graetz, J.J. Reilly, J. Johnson}
\affiliation{Department of Energy Sciences and Technology, Brookhaven National Laboratory, Upton, New York 11973}
\author{A.Y. Ignatov, T.A. Tyson}
\affiliation{Department of Physics, New Jersey Institute of Technology, Newark, New Jersey 07102}
\date{\today}
\baselineskip=10pt
\begin{abstract}
Ti $K$-edge x-ray absorption near edge spectroscopy (XANES) was used to explore the Ti valence and coordination in Ti-activated sodium alanate. An empirical relationship was
established between the Ti valence and the Ti $K$-edge onset based on a set of standards. This relationship was used to estimate oxidation states of the titanium catalyst in 2 mol\%
and 4 mol\% Ti-doped NaAlH$_4$. These results demonstrate that the formal titanium valence is zero in doped sodium alanate and nearly invariant during hydrogen cycling. A qualitative
comparison of the edge fine structure suggests that the Ti is present on the surface in the form of amorphous TiAl$_3$.
\end{abstract}
\maketitle
\clearpage

There is currently considerable interest in the development of metal hydrides capable of reversible hydrogen storage at low and medium temperatures.
A recent investigation of Ti-doped NaAlH$_4$ by Bogdanovic et al.~has demonstrated reversible cycling in the complex metal hydrides under
moderate conditions \cite{Bogdanovic1997}. Since this discovery there has been considerable work on improving the catalytic effects and
understanding the role of the catalyst in the sodium alanates. However, the mechanism by which the NaAlH$_4$ system is activated in the presence of a small amount of Ti is still not
well understood. In part, this is because the location and valence of the activating species is unknown.

In sodium aluminum hydride, the hydrogen absorption/desorption occurs through the following two-step decomposition-recombination
reaction:
\begin{equation}
\textup{NaAlH}_4 \leftrightarrow \frac{1}{3}\textup{Na}_3\textup{AlH}_6 + \frac{2}{3}\textup{Al} + \textup{H}_2 \leftrightarrow \textup{NaH} +
\textup{Al} + \frac{3}{2}\textup{H}_2
\label{reaction}
\end{equation}
giving a combined theoretical hydrogen storage capacity of 5.6 wt\%. There are a number of possible mechanisms by
which a metal dopant might enhance the dehydriding kinetics of equation \ref{reaction} \cite{Gross2000}. The simplest possibility is that the Ti acts as
a classic catalyst and assists the conversion of atomic hydrogen into molecular hydrogen at the surface. This situation is unlikely due to the strong
thermodynamic driving force on the Ti to form a hydride in the presence of the desorbed hydrogen.

Another proposed mechanism for the activation of NaAlH$_4$ involves the substitution of the dopant into the lattice. A recent x-ray diffraction study of
Ti and Zr-doped NaAlH$_4$ has demonstrated lattice distortions associated with doping the hydride \cite{Sun}. Based on these distortions, Sun et
al.~have suggested that the Ti is substituted for the Na cation \cite{Sun,Jensen1999} and is present as Ti$^{4+}$ at doping levels $\leq 2$ mol\%. The
enhanced kinetics are attributed to vacancy formation and the lattice distortion associated with the bulk lattice substitution.

A final possibility is that the hydrogen diffuses to the surface in an Al-H complex where it is dissociated by the catalyzing agent and releases H$_2$. This process leads to a
significant coalescence of Al on the particle surface in the dehydrided state. Energy dispersive spectroscopy of Ti-doped NaAlH$_4$ has revealed Al segregation towards regions of higher Ti concentration on the particle
surface upon the initial dehydrogenation \cite{Thomas2002}. The presence of zero-valent Ti is also supported by H$_2$ evolution during the doping process
\cite{Bogdanovic1997,Bogdanovic2000}. Other investigations of Ti-activated sodium alanate have demonstrated the formation of a TiAl$_3$ phase upon ball milling a
3:1 mixture of NaAlH$_4$ and TiCl$_3$
\cite{Majzoub2003}. However, it has been shown that only 2 \% of the metal is required to activate NaAlH$_4$ \cite{Jensen1999,Zidan1999,Sandrock2002}. Therefore, studies of
excessively-doped NaAlH$_4$ yield properties of the excess metal and not the activating species.

This letter presents XAS measurements of the Ti $K$-edge from the 2 mol\% and 4 mol\% Ti-doped NaAlH$_4$. The Ti valence was determined using the x-ray
absorption near-edge structure.

Ti-doped NaAlH$_4$ was prepared by mechanically milling 95\% pure NaAlH$_4$ (Alfa) with TiCl$_3$ (Aldrich 99\%) in a Fritsch Pulverisette 6
planetary mill. The powders ($\sim 1.2$ g) were milled in a 250 mL tungsten carbide bowl using seven 15 mm diameter WC
balls ($\sim 26$ g each). The mechanical attrition was performed in an inert Ar atmosphere for 1.5 hours. Samples containing 2 mol\% and
4 mol\% TiCl$_3$ were prepared for this study. The Ti-doped materials were hydrided/dehydrided in a calibrated fixed-volume for 4 cycles to ensure a
homogenous Ti distribution. TiH$_{1.1}$ was prepared by dehydriding TiH$_2$ (Alfa 99\%) at 700$^{\circ}$ C into a calibrated fixed volume. TiAl$_3$, Ti$_{0.08}$Al$_{0.92}$ and
Ti$_{0.02}$Al$_{0.98}$ were prepared by arc melting the components in a He atmosphere. The FeTi and TiAl$_3$ were crushed into a powder, while the more ductile alloys of
Ti$_{0.08}$Al$_{0.92}$ and Ti$_{0.02}$Al$_{0.98}$ were pressed into a thin pellet (123 MPa). Other Ti-containing materials included powders of TiO (Alfa 99.5 \%), Ti$_2$O$_3$ (Aldrich
99.9\%), TiO$_2$ (Johnson Matthey 99.995\%), FeTi (Canon-Muskegon 99\%), TiCl$_3$ (Aldrich 99.999\%) and TiH$_2$ (Alfa 99\%), and a foil of Ti metal (Materials Research Corp.
99.97\%). All powders were sieved through a 325 mesh (44 $\mu$m) and brushed onto Kapton tape.

Ti $K$-edge spectra were collected at beamline X-19A at the National Synchrotron Radiation Light Source (NSLS) using a Si(111) double crystal
monochromator. The higher-order harmonics were suppressed by detuning the second crystal of the monochromator on its rocking curve to 50 \% of the
maximum transmitted intensity at 400 eV above the edge. Spectra were recorded in fluorescence yield using a PIPS detector for the reasonably
concentrated samples and a 13-element Ge detector (Canberra) with an energy resolution of 240 eV for the diluted samples. All samples were kept at room temperature and oriented to
have the incident x-ray striking the surface at $45 \pm 5^{\circ}$. This geometry
provided essentially bulk-sensitive measurements. Due to small Ti concentrations, the XANES spectra were not corrected for "self-absorption". The energy
scale was calibrated by assigning $E=4966$ eV to the first inflection point of the pure Ti foil with an error of less than 0.1 eV. Air sensitive samples were measured in a sealed
sample holder after being loaded in a glove box under Ar. All spectra were background subtracted and normalized to a 100 eV window, 100 eV above the edge onset.

Powder x-ray diffraction (Rigaku Miniflex) of the Ti-doped NaAlH$_4$ after mechanical attrition revealed the presence of NaAlH$_4$ with
some minor concentrations of $\alpha$Na$_3$AlH$_6$, metallic Al, and NaCl. A small amount of NaCl is initially formed upon ball
milling, which results in the decomposition of a small amount of NaAlH$_4$. X-ray diffraction of the cycled samples suggests that additional NaCl (and Al
metal) is formed upon cycling. The majority species in both dehydrided samples were NaH and
metallic aluminum, consistent with equation \ref{reaction}. The hydrided 2 mol\% Ti sample consisted of predominately NaAlH$_4$, while the 4
mol\% Ti material exhibited more of a mixed phase, with NaAlH$_4$ and minor concentrations of $\alpha$Na$_3$AlH$_6$ and metallic Al, suggesting the
hydrogenation did not  go to completion in this sample. 
\begin{figure}
\includegraphics{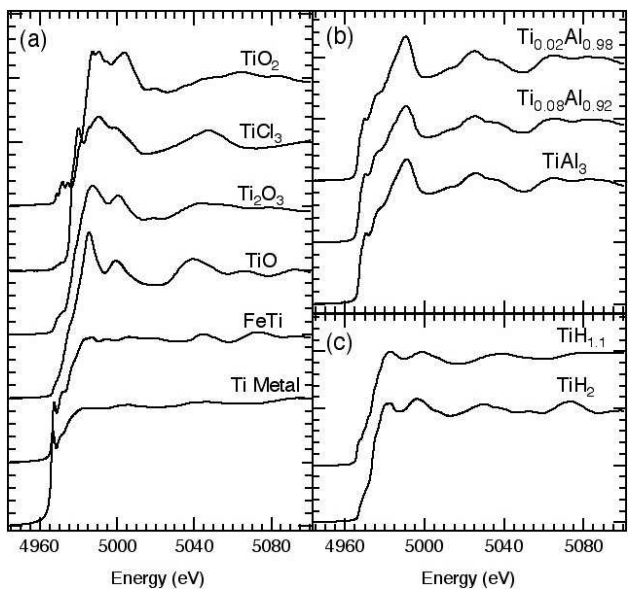} 
\caption{({\bf a}) Ti $K$-edges from Ti metal, FeTi, TiO, Ti$_2$O$_3$, TiCl$_3$, TiO$_2$, ({\bf b}) TiAl$_3$, nonstoichiometric Ti$_{0.08}$Al$_{0.92}$ and Ti$_{0.02}$Al$_{0.98}$,
({\bf c}) TiH$_2$ and nonstoichiometric TiH$_{1.14}$.}
\label{standards}
\end{figure}

The Ti $K$-edges from a series of Ti standards are displayed in Fig.~\ref{standards}a. The Ti edges for TiAl$_3$ and two Al samples with 2 mol\% and 8
mol\% Ti are shown in Fig.~\ref{standards}b. The edges from the three different Ti concentrations are qualitatively identical. This is not
surprising, since Ti is immiscible in Al at room temperature and will nucleate the TiAl$_3$ phase even at low dopant levels \cite{Massalski}. The Ti
$K$-edges from TiH$_2$ and TiH$_{1.1}$ (Fig.~\ref{standards}c) are also very similar, indicating that the structural environment around the Ti is essentially the
same. However,the local disorder in the substoichiometric compound (TiH$_{1.1}$) is clearly visible above the edge.
\begin{figure}
\includegraphics{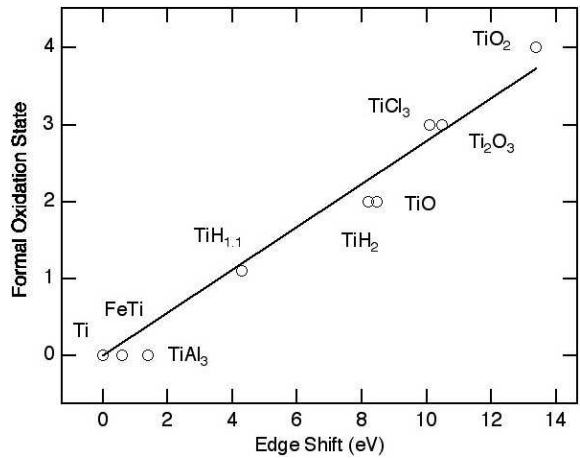} 
\caption{Shift in the Ti $K$-edge onset energy with respect to the formal oxidation state in a series of Ti standards.}
\label{oxidation_edgeshift}
\end{figure}

The most interesting features of Fig.~\ref{standards} are the large chemical shifts in the Ti $K$-edge, relative to Ti metal.  The shift of the Ti
$K$-edge reflects changes in the binding energy of the Ti 1$s$ electron. The largest observed shift is about 13 eV between Ti metal, containing neutral
Ti atoms, and TiO$_2$, containing Ti$^{4+}$. The increase in the $1s$ binding energy is due to the reduced screening of the nuclear
charge as the outer electrons are pulled off the atom.
\begin{figure}[b]
\includegraphics{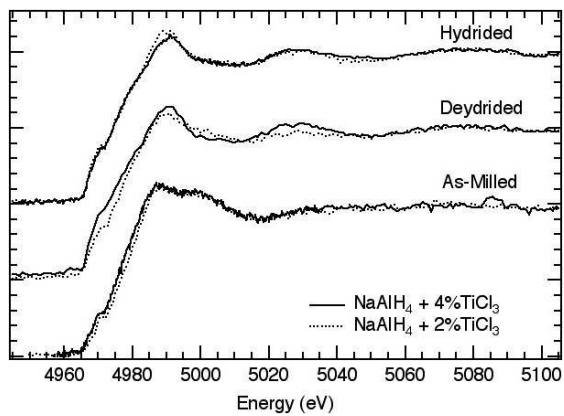} 
\caption{Ti $K$-edges from sodium alanate doped with 2 mol\% (thin line) and 4 mol\% Ti (thick line). Edges are shown for the
mechanically-milled material (no cycling), the dehydrided and the hydrided material after four cycles.}
\label{alanate_edges}
\end{figure}

A plot of the relationship between the edge onset and the Ti valence is shown in Fig.~\ref{oxidation_edgeshift}. The onset of the $K$-edge
was defined by the maximum of the first derivative. In cases where the peak in the first derivative was clearly split, such as TiH$_{1.1}$, the onset was chosen as the midpoint
between the two peaks. There is a linear correlation between the oxidation state of the Ti ion, $n$, and the edge shift with respect to Ti metal, $\Delta E$. This relationship ($n =
x\Delta E$) has a coefficient of $x = 0.28 \pm 0.01$.
\begin{table}
\caption{\label{tab:table1}Measured Ti $K$-edge onset energies and corresponding oxidation states for the Ti-doped sodium alanates}
\begin{ruledtabular}
\begin{tabular}{cccc}
Dopant & Condition & Edge Shift (eV) & Oxidation State \\
\hline
4 mol\% Ti & dehydrided & 0.6 & $0.17 \pm 0.01$\\
4 mol\% Ti & hydrided & 1.3 & $0.36 \pm 0.01$\\
2 mol\% Ti & dehydrided & 1.9 & $0.53 \pm 0.02$\\
2 mol\% Ti & hydrided & 1.9 & $\; 0.53 \pm 0.02$
\label{table}
\end{tabular}
\end{ruledtabular}
\end{table}

The edge onset energies were used to determine the Ti valence in the Ti-doped sodium alanates and
the results are displayed in Table \ref{table}. In the hydrided and dehydrided compounds the Ti is essentially zero-valent (total charge transfer of half
an electron, or less). Although the neutral character of the Ti valence is supported by a number of other studies
\cite{Bogdanovic1997,Bogdanovic2000,Balema2001,Thomas2002,Majzoub2003}, this is the first direct evidence of zero-valent Ti at dopant levels of
2--4 mol\%. The Ti $K$-edge from the as-milled material is believed to be a superposition of edges and therefore, no oxidation states were calculated. 

\begin{figure}
\includegraphics{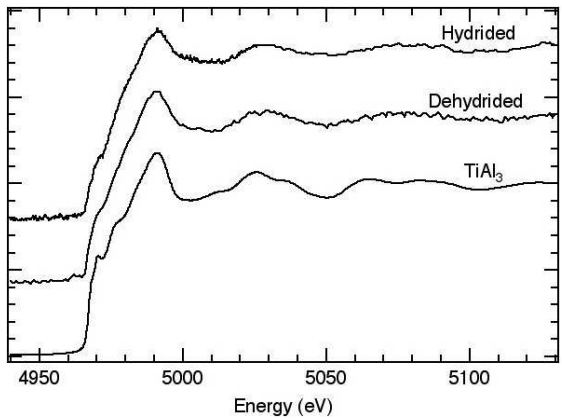} 
\caption{Ti $K$-edges from TiAl$_3$ and 2 mol\% Ti-doped sodium alanate in the hydrided and dehydrided states.}
\label{tiAl3_alanate}
\end{figure}

The structure of the Ti $K$-edge can also be used to gain insight into the local Ti environment. Figure \ref{tiAl3_alanate} displays a plot of the Ti $K$-edge from the hydrided and
dehydrided 2 mol\% Ti-doped sodium alanates and TiAl$_3$. In the doped sodium alanates, a small shoulder is evident just above the main peak (4998--5010 eV), which may be
attributed to the presence of a small amount of near-by hydrogen. The formation of titanium hydride is not unlikely since the presence of any Ti metal would likely form TiH$_{2-x}$ during
the first desorption. Despite this minor difference, the fine structure above the edge (4960--5100 eV) is qualitatively similar for the Ti-activated alanates and TiAl$_3$. This suggests
that the Ti is coordinated by Al atoms and that the local order is similar to that of TiAl$_3$. This is not surprising since TiAl$_3$ is the most
thermodynamically favorable Ti product ($\Delta H_{298 \textup{\scriptsize{K}}} = -142.4$ kJ/mol \cite{Smithells}), followed by TiH$_2$ ($\Delta H_{298 \textup{\scriptsize{K}}} = -119.7 $
kJ/mol \cite{crc64}) and Ti$_3$Al ($\Delta H_{298 \textup{\scriptsize{K}}} = -100.5$ kJ/mol \cite{Smithells}). However, the oscillations are clearly less pronounced in the doped alanates,
suggesting that the Ti lacks the long range order that exists in crystalline TiAl$_3$. This supports the recent study of Weidenthaler et al.~that demonstrated that Ti is atomically
dispersed and essentially amorphous in this system \cite{Weidenthaler}. The lack of long-range order would also explain why TiAl$_3$ has never been directly observed at these dopant
levels through diffraction techniques.

The Ti $K$-edge of the mechanically milled material (no cycling) exhibits peaks at 4988 eV 4998 eV (Fig.~\ref{alanate_edges}). This structure may be indicative of a superposition of edges
from Ti in multiple environments, the initial TiCl$_3$, TiH$_{2-x}$, and possibly TiAl$_3$. The decomposition of NaAlH$_4$ during mechanical milling liberates Al and
H$_2$, either of which may form a compound with Ti. Although TiAl$_3$ is more stable, the liberated H$_2$ is considerably more mobile and therefore more likely to react with Ti, forming
TiH$_{2-x}$. It is likely that the formation of TiAl$_3$ predominately occurs during the first decomposition, where metallic Al is more abundant. 

A final interesting feature of Fig.~\ref{alanate_edges} is the similarity of the hydrided and dehydrided spectra. This suggests that the local environment around the Ti is
nearly invariant during the hydrogen cycle (equation \ref{reaction}). Although there is no long-range order, the Ti is locally coordinated by Al at every stage of the
reaction.

These results confirm that the Ti is not present in the form of Ti metal and clearly demonstrate that there is no bulk lattice substitution. The catalyzing agent in Ti-doped
NaAlH$_4$ is present on the surface in the form of amorphous TiAl$_3$. 

This work was supported by the U.S. DOE contract DE-AC02-98CH10886 and NSF Contract DMR-0216858.

\bibliography{bibliography}
\bibliographystyle{unsrt}

\end{document}